\documentclass[12pt]{article}
\usepackage{isa_rmbs}
\usepackage{amsmath}
\usepackage{subfig}

\def\BibTeX{{\rm B\kern-.05em{\sc i\kern-.025em b}\kern-.08em
   T\kern-.1667em\lower.7ex\hbox{E}\kern-.125emX}}
\usepackage{wrapfig}				
\usepackage{graphicx}
\newcommand{\appropto}{\mathrel{\vcenter{
  \offinterlineskip\halign{\hfil$##$\cr
    \propto\cr\noalign{\kern2pt}\sim\cr\noalign{\kern-2pt}}}}}
\begin{document}

\title{Multiple Microtubule Tracking in Microscopy Time-Lapse Images Using Piecewise-stationary Multiple Motion Model Kalman Smoother}
\author{\textbf{S. Masoudi}\supit{a}, \textbf{C. H.G. Wright}\supit{a}, \textbf{N. Rahnavard}\supit{b}, \smallskip\\
            \textbf{J. C. Gatlin}\supit{c} and \textbf{J. S. Oakey}\supit{d}  \medskip\\
        \supit{a}Electrical and Computer Engineering Department\\ 
        University of Wyoming, Laramie, WY 82071 \smallskip\\
        \supit{b}Electrical and Computer Engineering Department\\ 
        University of Central Florida, Orlando, FL 32816 \smallskip\\
        \supit{c}Molecular Biology Department\\ 
        University of Wyoming, Laramie, WY 82071 \smallskip\\
        \supit{d}Chemical Engineering Department\\ 
        University of Wyoming, Laramie, WY 82071 \smallskip\\}

\date{} 
\maketitle
\begin{abstract}
Microtubules are inherently dynamic sub-cellular filamentuous polymers that are spatially organized within the cell by motor proteins which cross-link and move microtubules. \textit{In-vitro} microtubule motility assays, in which motors attached to a surface move microtubules along it, have been used traditionally to study motor function. However, the way in which microtubule-microtubule interactions affect microtubule movement remains largely unexplored. To address this question, time-lapse image series of \textit{in-vitro} microtubule motility assays were obtained using total internal reflection fluorescence (TIRF) microscopy. Categorized as a general problem of multiple object tracking (MOT), particular challenges arising in this project include low feature diversity, dynamic instability, sudden changes in microtubules’ motility patterns, as well as their instantaneous appearance/disappearance. This work describes a new application of piecewise-stationary multiple motion model Kalman smoother (PMMS) for modeling individual microtubules motility trends. To both evaluate the capability of this procedure and optimize its hyper-parameters, a large dataset simulating the series of time-lapse images was used first. Next, we applied it to the sequence of frames from the real data. Results of our analyses provide a quantitative description of microtubule velocity which, in turn, enumerates the occurrence of microtubule-microtubule interactions per frame.\\
\textbf{Keywords:} TIRF microscopy, velocity tracking, PMMS, microtubule-microtubule interaction.
\end{abstract}

 \section{Introduction}
Microtubules are the largest and stiffest type of the cytoskeleton polymer with two so-called plus and minus ends regarding their orientation inside the cell. Microtubules exist in the form of radial arrays, with their plus ends pointing outwards to the cell periphery during interphase~\cite{2,4,5,8,11}. This is how microtubules contribute to maintain the overall internal architecture of the cytoplasm \cite{6}. During mitosis, these radial arrays break down and build a machine, known as ``mitotic spindle"”, responsible for chromosome segregation during cell division~\cite{2, 6}. Additionally, microtubules participate in intra-flagellar cargo transport (IFT) and distribution of the organelles inside the cell as well as the cell motility~\cite{2,6,10}. Perturbation of normal microtubule function and arrangements is known to lead to a wide range of diseases from cancer to neurodegenerative disorders~\cite{pprezi}. Hence, having a quantitative analysis of microtubule dynamics seems critical to understand the underpinning mechanism of many diseases at the molecular level.

Generally speaking, microtubules exhibit two types of dynamics: (i) individually, each microtubule undergoes several stochastic transitions between its growth (subunit association) and shrinkage phases (subunit dissociation)~\cite{4,6,8,11}; (ii) microtubules interact with their surroundings, through direct contact or by crosslinking to other microtubules via microtubule associated proteins (MAPs). Thus far, various investigations have studied dynamic instability of the microtubules through tracking their plus-ends using microscopy-based time-lapse image sequences. However, the second form of their dynamics, i.e, changes in translation velocity due to interactions with other proteins and microtubules, is still an area that requires further inspection. Limited by other intracellular processes, we decided to employ \textit{in-vitro} microtubules motility assays, where the major contributor to their movements are the motor proteins attached to the coverslip. In doing so, we can assume the influence of other structures or processes is removed, reduced, or known~\cite{smal2010microtubule}.
While many studies have attempted to track individual microtubules through \textit{in-vitro} motility assays, the manner in which microtubule-microtubule interaction influences the microtubule movements/dynamics has not been their primary focus. Microtubules must interact with each other to fulfill several cellular functions, the most critical of which is chromosome segregation during mitosis. A subset of MAPs have inherent ATPase activity and function as motors which can walk along microtubules.   During mitotic spindle assembly, two motors are particularly important:  the minus-end directed motor cytoplasmic dynein (dynein) and the plus-end directed kinesin motor known as Eg-5.  These motors bind to, move, and spatially organize microtubules relative to one another~\cite{9,M1,M2,M3}. Image analysis used in this study should therefore inform us about microtubule velocities at each frame, elucidating the function of MAPs and their contribution to microtubules sliding, interaction, and assembly~\cite{M1,M2,M3}. 

For this work, microtubules assembled in Xenopus egg extracts spiked with fluorescently labeled tubulin heterodimers were recorded using time-lapse total internal reflection fluorescence (TIRF) microscopy~\cite{7}. Consequently, 2-D images of the fluorescently labeled microtubules were acquired sequentially~\cite{13}. Next, these images were processed to track individual microtubules, their trajectories were extracted, and their relevant motion parameters computed. Currently, manual analysis of such image data is the only available solution which is laborious, time consuming, heavily biased and often produces results with poor accuracy. Accordingly, there is a need for an automated method which works much faster and provides better accuracy based on an efficient low-cost algorithm.
  
 There are several challenges to be tackled while tracking multiple microtubules. First and foremost, is the low feature diversity of microtubules in terms of their shape and color. Second, these limited features may change over time, most  commonly as an outcome of dynamic instability which results in frequent changes of microtubules length. Third, the sudden appearance/disappearance of the microtubules as an artifact of using the TIRF microscopy (which illuminates a very thin layer immediate adjacent to the coverslip and provides the opportunity to observe moving microtubules only in a thin focal plane extending roughly $100$  \textit{nm} from the coverslip surface)~\cite{4,7}. Another challenge to tracking is the abrupt changes in microtubules movements, which may partly be caused by microtubule-microtubule interaction.

  The specific method we applied to solve the multiple microtubule tracking while addressing the above-mentioned issues, is explained in the Methods section. In order to evaluate and modify the processing algorithm, it was first applied to simulated data which imitates the actual microscopy time-lapse images of microtubules.  In the third section of this work, we provide the results of our tracking approach to simulated and real data. Later sections provide the relevant discussion and conclusions.
  
 During the last few decades, several speculations have been put forth to solve the general problem of multiple-particle tracking (MPT) in the context of sub-cellular particles where heterogeneous motion patterns are the major issue in this area, as is the case for microtubules~\cite{roudot2017piecewise}. Generally speaking, there are two steps in every tracking problem, 1) the recognition of relevant objects and 2) the linking step or measurement to prediction assignment. The latter has been addressed in literature using the advances in Bayesian filtering, multiple-hypothesis tracking methods, simulated annealing algorithm or batch approaches. While all these techniques try to provide better temporal consistency by taking into account the past and/or future time points of a fixed temporal window, disregarding the motion model, there are motion modeling approaches in contrast which mostly use standard interacting multiple model (IMM) filtering to assign a cost to each possible track based on past information only~\cite{roudot2017piecewise}. Consequently, a high cost is attributed to tracks presenting unpredictable changes of motion type, leading to less chance for them to be selected by the optimization algorithm. According to recent reviews, the multi-frame tracking based algorithms are useful in scenarios with numerous false positives while the motion modeling approaches outperform other methods in cases with motion model complexity~\cite{smal2015quantitative}. In the work recently published by Roudot et al., these two different perspectives are combined in as a piecewise-stationary motion model (PMM) for the particle transport along an iterative smoother that exploits recursive tracking in multiple rounds in both forward and backward temporal directions \cite{roudot2017piecewise}.
 
 \section{Methods}
 \paragraph{Kalman Filter}
 Linearity between consecutive time points of a particle movement in its state space form, where both measurement noise and model noise follow normal distributions, can allow use of the Kalman filter to optimally estimate its state \cite{roudot2017piecewise}.
 \begin{align}
 \nonumber &\textbf{x}_{t+1}=\textbf{F}\textbf{x}_{t}+\textbf{W}_{t+1}\\
 &\textbf{z}_{t+1}=\textbf{H}\textbf{x}_{t+1}+\textbf{v}_{t+1}
  \label{statespace}
 \end{align}
  Usually, in order to track a single particle at time $t$, the state vector $\textbf{x}_{t}$ includes both position $(x_{t},y_{t})$ and speed $(dx_{t},dy_{t})$ in two dimensions while the transition matrix $\textbf{F}$ and the observation matrix $\textbf{H}$ are defined as follows respectively to express the constant velocity or directed motion model type and to project the state from state space to the measurement space \cite{roudot2017piecewise}:
 \begin{equation}
 \textbf{x}_{t}=(x_{t},y_{t},dx_{t},dy_{t})^{T}\qquad \textbf{F}=
 \begin{bmatrix} 1& 0 & 1 & 0 \\ 0& 1 & 0 & 1 \\0& 0& 1 &0 \\0& 0 & 0 & 1  \end{bmatrix} \qquad \textbf{H}=
 \begin{bmatrix} 1& 1 & 0 & 0  \end{bmatrix}
 \end{equation}
 There are also $\textbf{w}_{t+1}$ and $\textbf{v}_{t+1}$ that represent the white noise in model and measurement equations with covariance matrices $\textbf{Q}_{t}$ and $\textbf{R}_{t}$. According to the Bayesian framework, the probability distribution of $\textbf{x}_{t}$ can be sequentially estimated based on the provided observation. For filtering, two episodes of prediction and update are expressed as:  
 \begin{equation}
 \begin{aligned}[c]
 \bar{\textbf{x}}_{t+1}&=\textbf{F}\hat{x}_{t}\\
 \bar{\textbf{P}}_{t+1}&=\textbf{F}\hat{P}_{t}\textbf{F}^{T+}\textbf{Q}_{t}
 \end{aligned}
 \qquad
  \begin{aligned}[c]
 \textbf{K}_{t+1}&=\frac{\bar{\textbf{P}}_{t+1}\textbf{H}^{T}}{\textbf{H}\bar{\textbf{P}}_{t+1}\textbf{H}^{T}+\textbf{R}_{t+1}}\\
  \hat{\textbf{P}}_{t+1}&=(\textbf{I}-\textbf{K}_{t+1}\textbf{H})\bar{\textbf{P}}_{t+1}\\
  \hat{\textbf{x}}_{t+1}&=\bar{\textbf{x}}_{t+1}+(\textbf{z}_{t+1}-\textbf{H}\bar{\textbf{x}}_{t+1})\textbf{K}_{t+1}
 \end{aligned}
\label{Eq_Correction}
 \end{equation}
in which $\textbf{K}_{t+1}$ represents the adaptive Kalman gain used to limit the measured innovation along with the bar and hat signs denote the predicted and updated values, respectively \cite{roudot2017piecewise}. In case of tracking multiple particles, there will be an additional assignment step right after the prediction and before the updating phase, to individually assign various predictions from the set of tracks to their best matching measurements at time $t+1$. Later, during the update step, the predictions will be modified using new available measurements\cite{roudot2017piecewise}.
 
\paragraph{Multiple motion modeling}
Assigning a single Kalman filter to each object in MPT, cannot take into account the common motion heterogeneity in intracellular transport \cite{roudot2017piecewise}. While upgrading from a single constant speed model to more complicated models may seem effective, they are still insufficient to react to unpredictable changes in the observed motion of a particle. To handle multiple motion types for each object, papers in the literature propose using multiple Kalman Filters:
 \begin{align}
\nonumber &\textbf{x}_{t+1}=\textbf{F}^{\theta_{t+1}}\textbf{x}_{t}+\textbf{W}_{t+1}^{\theta_{t+1}}\\
 &\textbf{z}_{t+1}=\textbf{H}^{\theta_{t+1}}\textbf{x}_{t+1}+\textbf{v}_{t+1}^{\theta_{t+1}}
 \label{Sysmod}
 \end{align}
where indexed notation $\theta_t$ stands for the model type at time $t$ which may be any of $N$ predefined models \cite{roudot2017piecewise}. Accordingly, to obtain the pdf of the state $\textbf{x}_{t+1}$, all possible model type sequences $\theta_{1:t}$ should be considered. However, the parameter space describing this sequence increases exponentially with time, which in turn needs an exponentially growing number of Kalman filters to estimate the \textit{a posteriori} distribution of each mode sequence. To approximately estimate this pdf instead, possible models at time $t+1$ are only considered. 
 \begin{align}
 \nonumber p(\textbf{x}_{t+1},\textbf{z}_{1:t+1})&\approx\sum_{\theta_{t+1}}{p(\textbf{x}_{t+1},\textbf{z}_{1:t+1},\theta_{t+1})} 
 \\
 &\appropto \sum_{\theta_{t+1}}{p(\theta_{t+1}|\textbf{z}_{1:t+1})p(\textbf{x}_{t+1}|\textbf{z}_{1:t+1},\theta_{t+1})}
 \end{align}
 which expresses a general Gaussian mixture at time step $t+1$, considering $p(\textbf{x}_{t+1}|\textbf{z}_{1:t+1},\theta_{t+1})$ is a Gaussian distribution following $N(\hat{\textbf{x}}_{t+1}^{\theta_{t+1}},\hat{\textbf{P}}_{t+1}^{\theta_{t+1}})$\cite{roudot2017piecewise}. In fact, different implementations of multiple motion models are built on different choice or modeling of $p(\textbf{x}_{t+1}|\textbf{z}_{1:t+1},\theta_{t+1})$. Here we applied the piecewise-stationary motion model filtering (PMM).
\paragraph{Piecewise-stationary motion model filtering}
Using the model proposed in \cite{roudot2017piecewise}, for the system described by Equation \ref{Sysmod}, we assumed $M$ successive stationary regimes of $\{\theta_0,...,\theta_{M-1}\}$ along the time with starting/ending time points to be $\{T_0,...,T_{M}\}$. Thus we will have:
\begin{align}
 p(\textbf{x}_{t+1},\textbf{z}_{1:t+1})&= p(\textbf{x}_{t+1},\textbf{z}_{T_i:t+1})\nonumber\\
 &\approx\sum_{\theta_{i}}{p(\theta_{i}|\textbf{z}_{T_i:t+1})p(\textbf{x}_{t+1}|\textbf{z}_{T_i:t+1},\theta_{i})}
 \end{align}
where $T_{i}$ is the starting time of the $i^{th}$ motion regime $\theta_{i}$ (current). The conditional state probability will be derived through a Kalman filter prediction and update phases based on a mechanism known as ``static multiple modeling" with the filtering parameters as \cite{roudot2017piecewise}:
\begin{equation}
\begin{aligned}[c]
 \bar{\textbf{x}}^{\theta_{i}}_{t+1}&= {\textbf{F}}^{\theta_{i}}\hat{\textbf{x}}^{\theta_{i}}_{t+1}\\
 \bar{\textbf{P}}^{\theta_{i}}_{t+1}&= {\textbf{F}}^{\theta_{i}}\hat{\textbf{P}}^{\theta_{i}}_{t+1}{{\textbf{F}}^{\theta_{i}}}^{T}+{\textbf{Q}}^{\theta_{i}}_{t}\\
  \end{aligned}
  \qquad
  \begin{aligned}[c]
 \textbf{K}^{\theta_{i}}_{t+1}&=\frac{\bar{\textbf{P}}^{\theta_{i}}_{t+1}{\textbf{H}^{\theta_{i}}}^{T}}{\textbf{H}^{\theta_{i}}\bar{\textbf{P}}^{\theta_{i}}_{t+1}{\textbf{H}^{\theta_{i}}}^{T}+\textbf{R}^{\theta_{i}}_{t+1}}\\
  \hat{\textbf{P}}^{\theta_{i}}_{t+1}&=(\textbf{I}-\textbf{K}^{\theta_{i}}_{t+1}\textbf{H}^{\theta_{i}})\bar{\textbf{P}}^{\theta_{i}}_{t+1}\\
  \hat{\textbf{x}}^{\theta_{i}}_{t+1}&=\bar{\textbf{x}}^{\theta_{i}}_{t+1}+(\textbf{z}_{t+1}-\textbf{H}^{\theta_{i}}\bar{\textbf{x}}^{\theta_{i}}_{t+1})\textbf{K}^{\theta_{i}}_{t+1}
 \end{aligned}
 \label{Filtering}
 \end{equation}
\paragraph{Piecewise-stationary multiple motion model Kalman smoother(PMMS)}
To maximize the amount of information, a Kalman smoother is used which exploits both past and future measurements at the same time\cite{roudot2017piecewise}. The approach is to integrate two independent Kalman filters carried out in forward and backward temporal orders into a smoother \cite{roudot2017piecewise}. Hence, as specified by \cite{kitagawa1994two}, $p(\textbf{x}_{t}|\textbf{z}_{1:T})$ follows a Gaussian distribution with the mean and covariance formulated in terms of the mean and covariance in forward and backward directions as in Equation \ref{eq87} .
\begin{equation}
\begin{aligned}[c]
\hat{x}_{t}=\hat{P}_{t}({\hat{P}_{t}}^{f^{-1}}{\hat{\textbf{x}}_{t}}^{f}+{\hat{P}_{t}}^{b^{-1}}{\hat{\textbf{x}}_{t}}^{b})
 \end{aligned}
 \qquad
 \begin{aligned}[c]
\hat{P}_{t}=({\hat{P}_{t}}^{f^{-1}}+{\hat{P}_{t}}^{b^{-1}})^{-1}
 \end{aligned}
 \label{eq87}
 \end{equation}
 \section{Results}
\begin{figure}[!ht] 
 \advance\leftskip-1cm
 \subfloat[]{\includegraphics[width=0.56\textwidth,height=0.27\textheight]{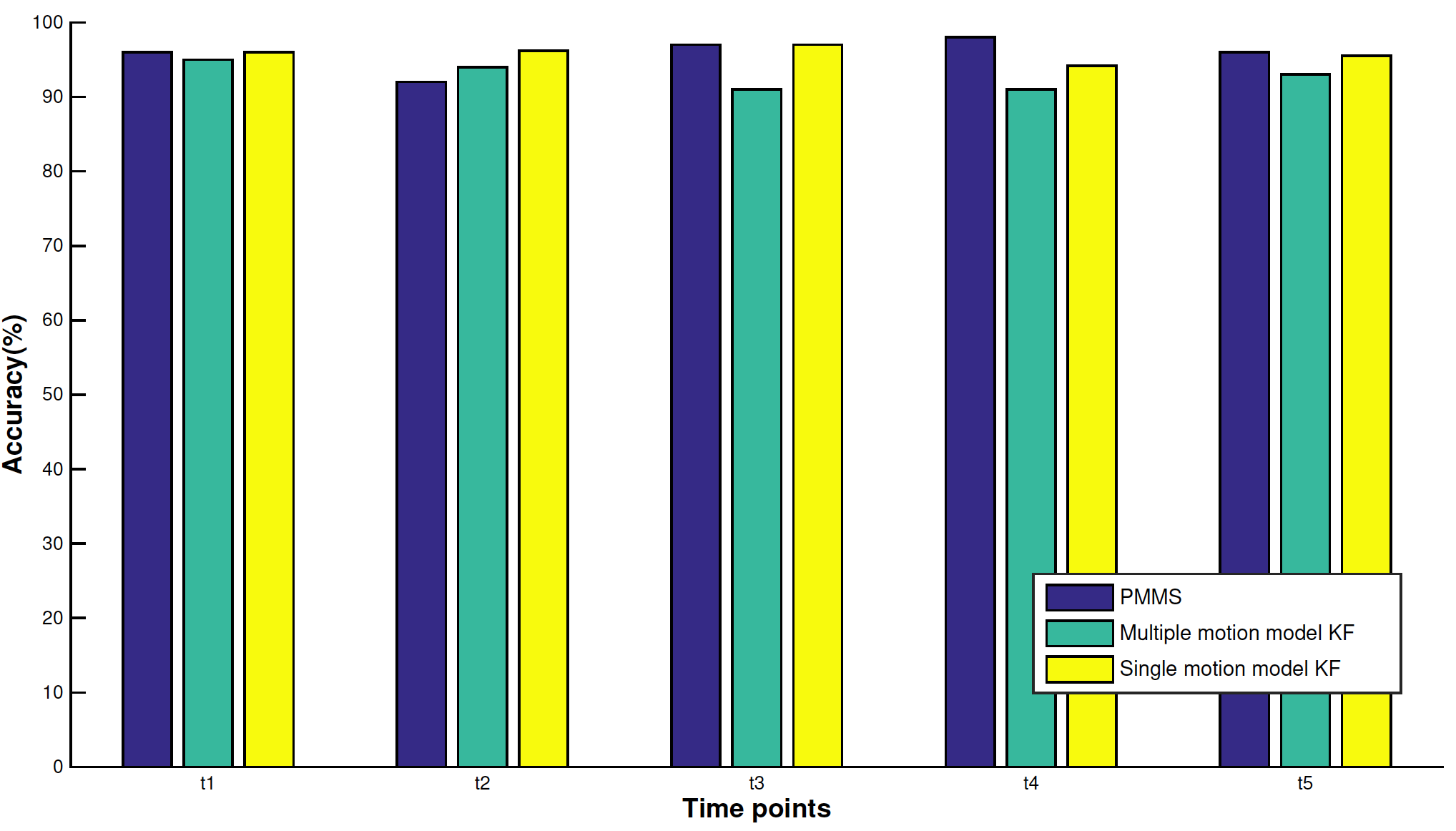}} 
 \subfloat[]
{\includegraphics[width=0.56\textwidth,height=0.27\textheight]{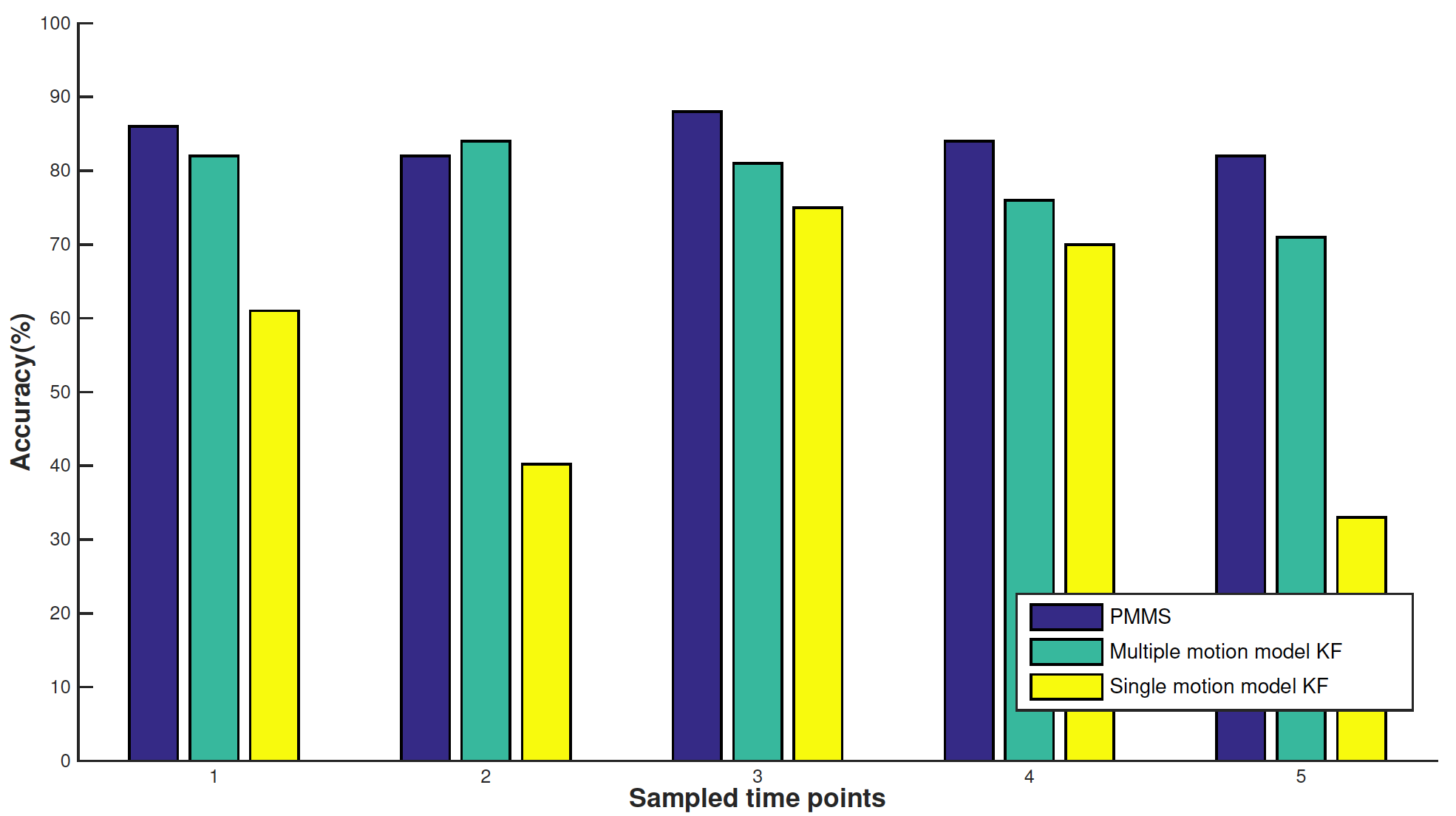}} 
  \caption{Velocity estimation results, averaged over $15$ microtubules along $5$ sampled time-points (a) freely moving with no interaction (b) possibly interacting.} 
  \label{Plot2}
\end{figure}

 \paragraph{Simulated data}
To examine the accuracy of particle detection and trajectory estimation, we first applied the PMMS method to the simulated data and compared its results to the multiple model and single model Kalman filters. The simulated motion in this data is dominated by several rules concerning the initial condition, the probability distribution of the microtubules numbers, appearance/disappearance, dynamic instability and motion variability. Results are presented in two forms of the accuracy via motion type switching probability and accuracy via microtubule numbers. Results definitely confirm the superior behavior of the PMMS algorithm over two other approaches. 
\begin{figure}[!ht] 
\advance\leftskip-1cm
\subfloat[]{\includegraphics[width=0.57\textwidth,height=0.27\textheight]{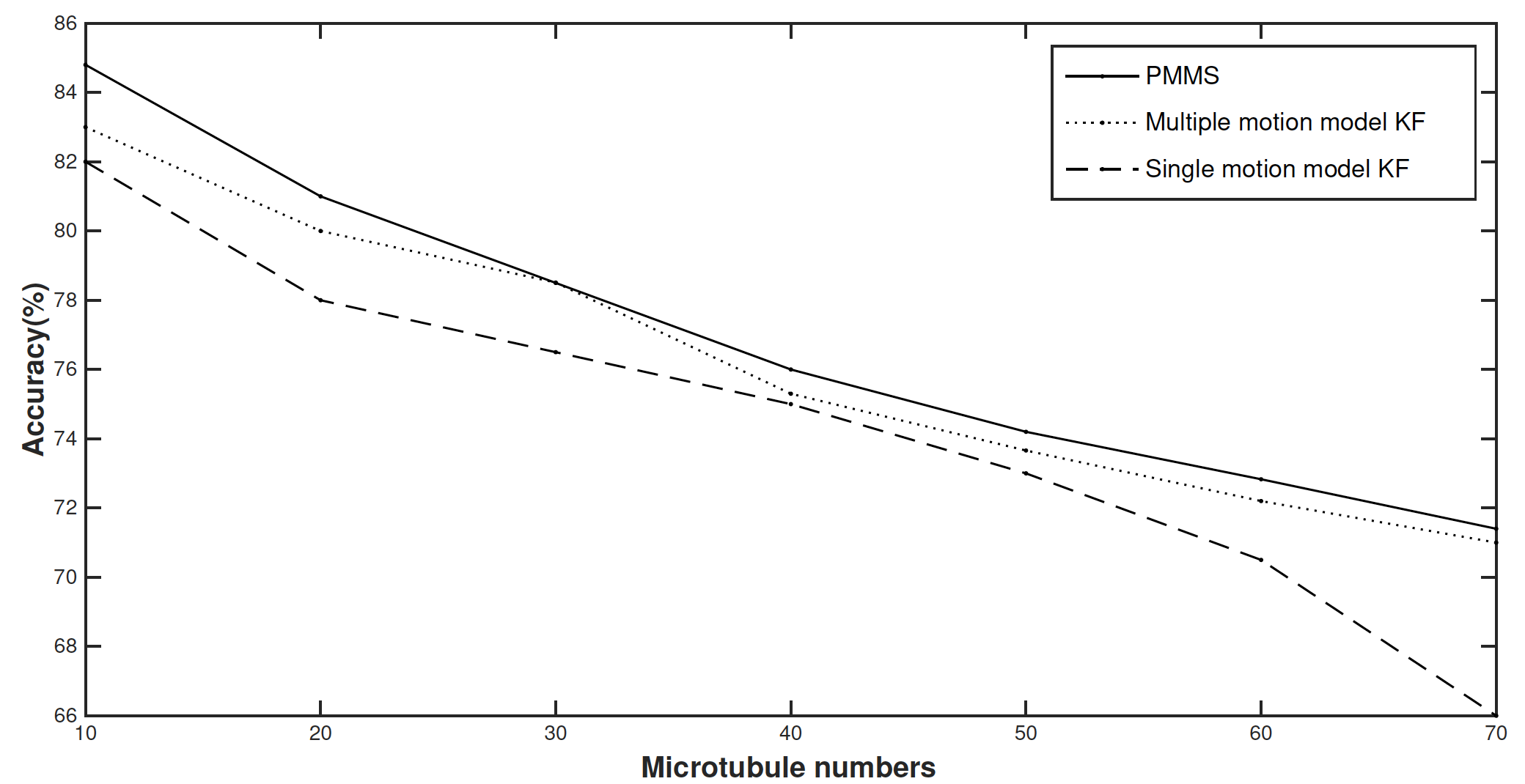}}
\subfloat[]{\includegraphics[width=0.57\textwidth,height=0.27\textheight]{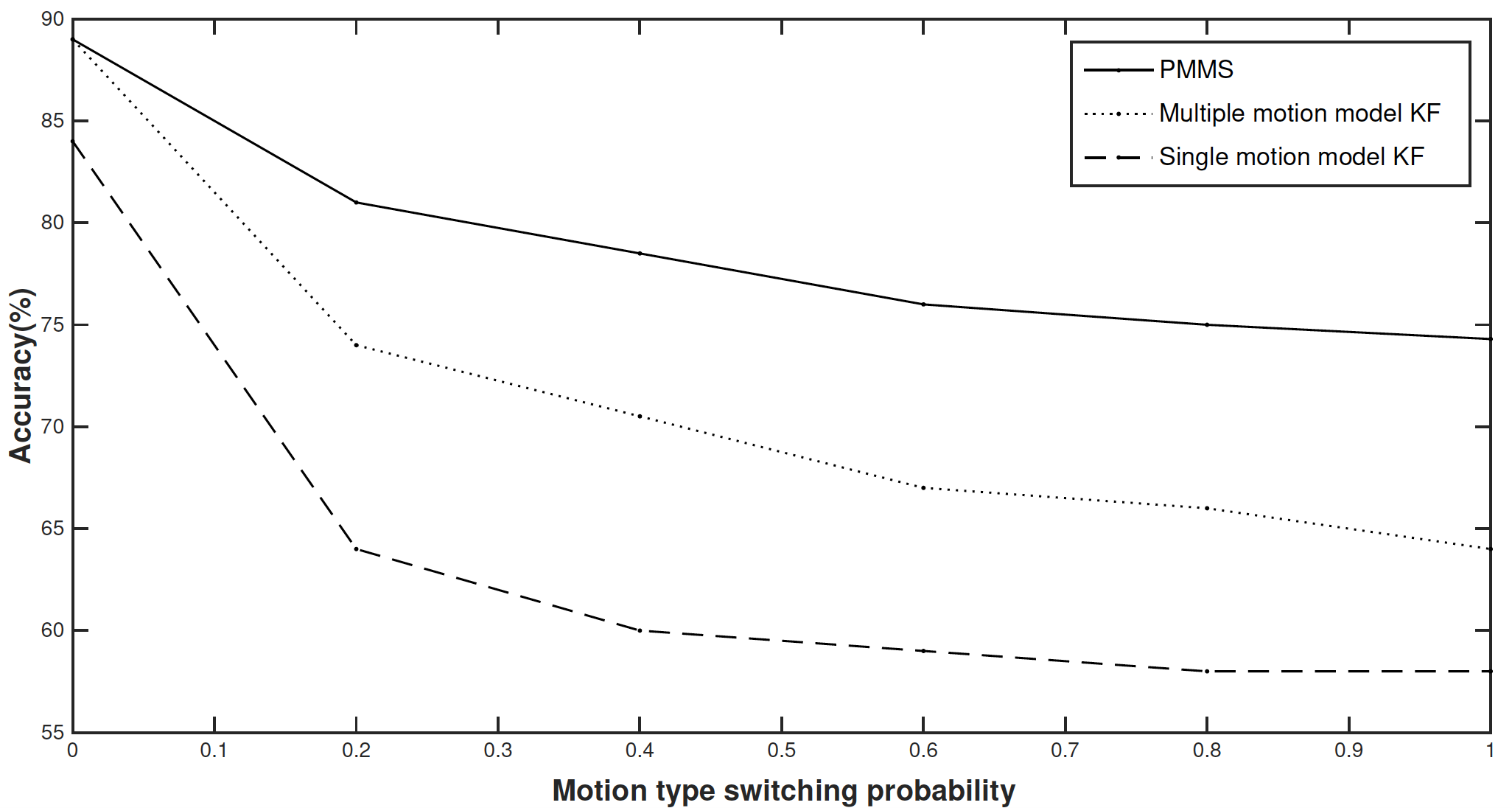}} 
\caption{True positive ratio results obtained using three methods (a) on the same simulated data with initially 50 microtubules (b) on various simulated data with different number of microtubules and motion type switching probability of 0.3. } 
  \label{Plot1}
\end{figure}
 \paragraph{Real Data}
The results for velocity estimation in form of $1-\frac{|{\textbf{annotated value-estimated value}}|}{\textbf{annotated value}}$ are shown in Figure \ref{Plot2}.  Figure \ref{Plot2}-(a) is based on the averaged velocity of $15$ free individual microtubules along $5$ sampled time-points, i.e, these microtubules are not interacting with any other. It can be seen that the single motion model Kalman filter works just as well as the other two methods, since microtubules move in a unique motion type. Same results in case of possibly interacted microtubules are presented in Figure \ref{Plot2}-(b). In this case, $5$ time-points are sampled in an ordered manner so that they represent before, during, and after the time interval of microtubules passing by. A dramatic difference in results of the three algorithms can be seen in Figure \ref{Plot2}-(b) where microtubules go through a motion transition caused by their possible interaction with each other. This difference is magnified while (i) passing from $t_1$ representing the time before possible interaction to $t_2$ during which possible interaction happens, (ii) when passing from $t_4$ to $t_5$ which illustrates the transition from interaction to free movement.

\vspace{-4mm}
\section{Discussion}
Confirmed by the obtained results, the Bayesian framework in general shows a great capability in dealing with complicated problem of microtubules' velocity estimation. It can be seen that applying an algorithm with a more characteristic model and wider window of observations along time, makes it more complicated, which enables us to obtain better results. A next line of research would be to take the concept of the particle filter into consideration, which can be used to form a multiple motion model particle smoother. Although equipping the algorithms to successfully track microtubules and estimate their velocities in confusing scenarios seems to be a great solution, there still may be a need to achieve better results using a general solution not limited to manual selection of features. For this purpose, convolutional neural networks may be a powerful substitute for which the features of interest are trained to be picked by the algorithm itself.

\vspace{-4mm}
\section{Conclusions}
 In order to detect microtubule-microtubule interactions, several methods have been applied to track their velocity. The compatible framework of these methods, with the probabilistic  nature of microtubules motion, leads to reasonable results. The complicated implementation of these algorithms, however, may be a disadvantage that makes them impractical in the case of on-line tracking. Even so, the significance of the obtained results shows the potential for automatic off-line processing of time-lapse microscopy images of microtubules. As this research progresses, it is hoped that it will provide a better understanding of the underlying mechanics of cancer.

 \bibliographystyle{ieeetr} 

\end{document}